\documentclass[noshowpacs,amsmath,twocolumn,8pt,aps,prb]{revtex4}  % this article uses the Revtex 4-1 format
\usepackage{setspace}
\usepackage{amsmath}
\usepackage{graphicx,afterpage}
\usepackage{verbatim}  % for commenting
\usepackage{amsfonts}
\usepackage{amssymb}

\begin{document}

\title{Chip-based Brillouin lasers as spectral purifiers for photonic systems}         % Enter your title between curly braces
\author{Jiang Li, Hansuek Lee, Tong Chen, Oskar Painter and Kerry Vahala }
\affiliation{T. J. Watson Laboratory of Applied Physics, California Institute of Technology, Pasadena, California 91125, USA}
\date{\today}

\maketitle

\textbf{High coherence lasers are essential in a wide range of applications, however, such performance is normally associated with large laser cavities, because increasing energy storage reduces quantum phase noise and also renders the laser frequency less sensitive to cavity vibration. This basic scaling rule is at odds with an emerging set of optical systems that place focus on compact (optimally integrable) sources of high coherence light.  These include phase-coherent optical communication using quadrature-amplitude-modulation, and also record-low phase noise microwave sources based upon optical comb techniques.  In this work, the first, chip-based Brillouin laser is demonstrated.  It features high-efficiency and single-line operation with the smallest recorded Schawlow-Townes frequency noise for any chip-based laser. Because the frequency offset between the laser's emission and the input pump is relatively small, the device provides a new function: spectral purification of compact, low coherence sources such as semiconductor lasers. }

Ultra-high coherence (low frequency noise) has emerged as a priority in a remarkably wide range of applications including: High-performance microwave oscillators \cite{Diddams}, coherent fiber-optic communications \cite{coherentcom1}, remote sensing \cite{Lidar} and atomic physics \cite{Bergquist, bergquist2}.  In these applications, the laser  forms one element of an overall system that would benefit from miniaturization.  For example, in coherent communication systems a laser local oscillator works together with taps, splitters and detectors to demodulate information encoded in the field amplitude of another coherent laser source.  A requirement in these systems is to create narrow-linewidth lasers so that a large number of information channels can be encoded as constellations in the complex plane of the field amplitude \cite{coherentcom1, coherentcom2}.  In yet another example, the lowest close-to-carrier phase-noise  microwave signals are now derived through frequency division of a high-coherence laser source using an optical comb as the frequency divider \cite{Diddams}.  The miniaturization of these all-optical microwave oscillators could provide a chip-based alternative to electrical-based  microwave oscillators, but with unparalleled phase noise stability. Moreover, with the advent of microcombs \cite{TJK1,TJKreview}, such an outcome seems likely provided that similar strides are possible in miniaturization of high-coherence sources and reference cavities.  

\begin{figure}[h]
\begin{center}
\includegraphics[width=0.5\textwidth]{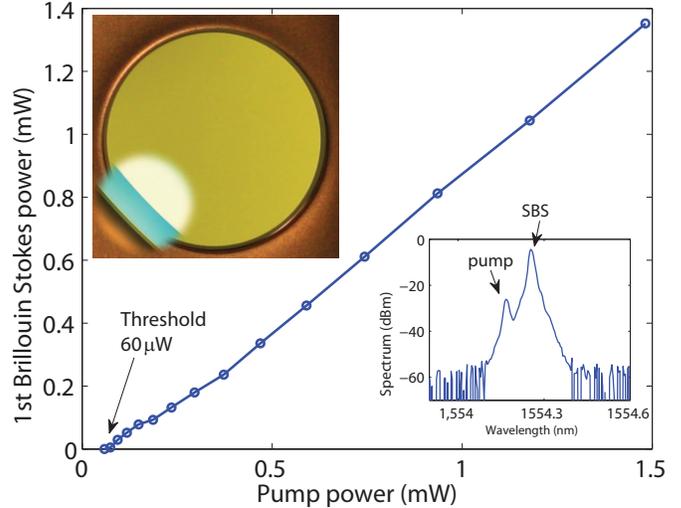}
\caption{\label{fig:LLcurve}\textbf{Data showing output SBL power versus input pump power}. Insets: (upper) Image of SBL ultra-high-Q resonator with magnified edge region; (lower) optical spectrum of the SBL with the pump suppressed  on account of the contra-directional emission of the SBL.}
\end{center}
\end{figure}

Typically, however, miniaturization comes at the expense of coherence, because quantum and technical noise contributions to laser coherence increase as laser-cavity form factor is decreased. In this work, we demonstrate a stimulated Brillouin laser that attains the highest level of coherence reported for a chip-based device. Schawlow-Townes noise less than 60 milliHz$^{2}$/Hz is measured at an output power of approximately 400$\mu$W. Also significant is that the low-frequency technical noise is comparable to commercial fiber lasers.  The devices use a new ultra-high-Q cavity \cite{Disk FIO, Disk axiv} that enables precise matching of the Brillouin gain shift to the free-spectral-range, thereby guaranteeing reliable oscillation  (see inset in Figure \ref{fig:LLcurve}). The devices are efficient, featuring more than 90\% conversion of lower-coherence pump to high-coherence output and threshold powers as low as 60 microWatts.  

A pre-requisite for high coherence is that the oscillator must operate single line  with high sidemode suppression.  Mutli-line operation not only lowers the overall coherence of source, but the presence of even relatively weak sidemodes introduces low-frequency, mode partition noise \cite{agrawal}. There are also significant collateral problems caused by multiline operation that are specific to certain applications (e.g.,  dispersion in optical fiber communication).  Intracavity etalons or grating filters are normally used to induce high levels of mode selection in lasers.  In the present work, the relatively narrow SBS gain spectrum provides high mode selection without the need for etalons or grating filters.

Single-line oscillators suffer from two general types of laser frequency noise: one that is associated with technical noise such as from thermal drift and vibration; and a second of fundamental origin, the Schawlow-Townes (ST) noise from spontaneous emission.  ST noise is broadband and has a double-sided frequency-fluctuation spectral density given by\cite{STownes, Vahala}:

\begin{equation} 
\label{eq:STnoise}
S_{\nu}^{ST}(f)=\frac{ \mu h \nu^{3}}{2P Q_{T}Q_{ex}}
\end{equation}

{\noindent}where $\mu$ is the spontaneous emission factor, h is the Planck constant, $\nu $ is the laser frequency and P is the output power; $Q_{T}$ is the total cavity quality factor and relates to the external cavity quality factor $Q_{ex}$ and intrinsic cavity quality factor $Q_{0}$  through  $\frac{1}{Q_{T}}=\frac{1}{Q_{ex}}+\frac{1}{Q_{0}}$. In an ideal case of no other contribution to frequency noise, the spectral lineshape of the laser is Lorentzian with a full-width-half-maximum linewidth given by $\Delta \nu = 2\pi S_{\nu}^{ST}(f)  $. The ST noise scales inversely with laser output power and inverse quadratically with the laser Q factor \cite{STownes, Vahala}. This latter dependence is responsible for the difficulty in maintaining high coherence as device form factor is reduced; and nearly all laser cavities suffer Q factor degradation as the cavity size is reduced, either on account of increasing output coupling losses as in a Fabry-Perot laser or on account of the challenge presented by fabrication of high-Q micro-cavities in a chip-based device.  For example, state-of-the-art, high-coherence, monolithic semiconductor lasers feature an ST linewidth in the range of 3 kHz \cite{semiconductor} by using a long-cavity, corrugation-pitch-modulated laser structure.  While these levels are currently acceptable for coherent communications, further enhancements in coherence would improve performance. Morever, other applications, such as to microwave oscillators, require nearly 4 orders of magnitude narrower linewidths. 

Among optically pumped,  chip-based laser oscillators, a beat note linewidth in a split-mode,  Erbium-doped microtoroid laser produced a value of 4 Hz \cite{sogel} and a microtoroid Raman laser showed an ST linewidth of 3 Hz \cite{Raman}.  However, the Raman and Erbium gain spectra are very broad so that no frequency selection mechanism exists in these devices. As such, the devices are not well controlled spectrally and frequently lase on multiple lines or exhibit mode hopping \cite{Raman Cascade}.  Recently, a related process, stimulated Brillouin laser (SBL) action, has been demonstrated in discrete micosphere \cite{Tal} and CaF$_{2}$ micocavities \cite{Ivan}. While Brillouin scattering is well known in optical fiber communication, including its use for gain in narrow-linewidth fiber lasers \cite{fiberSBS1}, generation of slow light \cite{Geata, Allan} and information storage \cite{Boyd}, realization of microcavity based SBLs has proven very challenging on account of the requirement to precisely match the Brillouin shift to a pair of cavity modes. Specifically, the narrow linewidth of the Brillouin gain requires better than 1:1000 control of the resonator diameter to obtain a match and more realistically it requires 1:10,000 control for consistent low-threshold turn-on power. 

In the recent demonstrations, an approach was implemented in which the high spectral density of transverse modes in spheres or CaF$_{2}$ resonators offered a reasonable likelihood of finding a well-matched pair of resonator modes. In the present work, this approach is replaced by precision control of the cavity free spectral range (FSR) using a new ultra-high-Q resonator geometry \cite{Disk FIO, Disk axiv}. These devices are the first chip-based Brillouin devices.  Their coherence sets a record for frequency stability of any chip-based device. The technical noise of the devices is also remarkably low and comparable to commercial fiber lasers (currently a benchmark for good frequency stability). Moreover, the same issue that has made these devices so difficult to fabricate  in microcavity form(the relatively narrow gain spectrum) becomes an asset in creating highly stable, single-line oscillation.  As now shown, the relatively small offset frequency between the pump and the laser output of these devices makes these devices well suited as compact spectral purifiers in any system requiring high coherence.

\begin{figure}[h]
\begin{center}
\scalebox{0.38}{\includegraphics{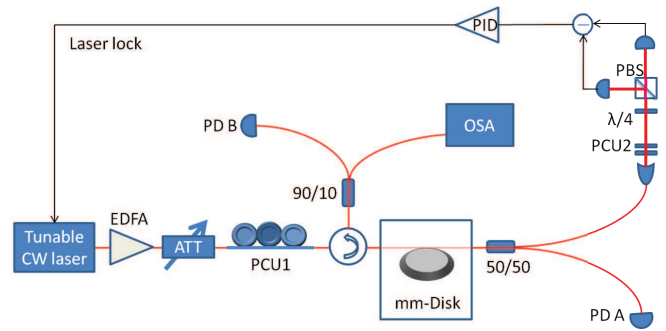}}
\caption{\label{fig:schematic}\textbf{Experimental setup.} A tunable CW laser is amplified through an EDFA and coupled into the disk resonator using the taper-fiber technique. The SBL propagating in the backward direction propagates through the fiber circulator, and is monitored by photodetector (PD B) and optical spectrum analyzer (OSA). The pump is monitored by a separate photodetector (PD A) and is also coupled to a balanced-homodyne detection setup (H\"{a}nsch-Couillaud technique) to generate an error signal for locking the pump laser to the cavity resonance.}
\end{center}
\end{figure}

The resonators used in this work are new and attain Q factors as high as 875 million \cite{Disk FIO, Disk axiv}, even exceeding the performance of microtoroids \cite{toroid}. Significantly, these devices are fabricated entirely from standard lithography and etching, avoiding any kind of silica reflow step \cite{toroid}. The combination of ultra-high-Q and precision control of FSR has not previously been possible, and is essential for reliable fabrication of low-threshold SBL lasers. Very briefly, devices use an 8-10 micron thick thermal oxide layer that is lithographically patterned and etched using buffered HF into disks. The etched, oxide disks then act as a mask for selective dry etching of the silicon. This dry etch process creates a whispering gallery resonator through undercut of the silicon. By proper control of both the wet and dry etching processes, the Q of the resulting resonator can be nearly 1 billion. Moreover, the lithography and etching process provide diameter control of 1:20,000, more than sufficient to place the microcavity FSR within the Brillouin frequency shift. The upper left inset in Figure~\ref{fig:LLcurve} shows a top view of a resonator fabricated using this procedure. Additional details on fabrication and characterization of these resonators are given elsewhere\cite{Disk FIO, Disk axiv}. 

Figure ~\ref{fig:schematic} shows the experimental setup used to test the SBLs. Pump power is coupled into the resonator by way of a fiber taper coupler \cite{Cai,ideality}. SBL emission is coupled into the opposing direction and routed via a circulator into a photodiode, optical spectrum analyzer or an interferometer (not shown) for measurement of frequency noise. The transmitted pump wave is monitored using a balanced homodyne receiver so as to implement a H\"{a}nsch-Couillaud  locking of the pump to the resonator \cite{HC,TJK3}.

By proper control of taper loading, the SBL can be operated in two distinct ways: cascade or single-line. In cascade, the waveguide loading is kept low so that once oscillation on the first Stokes line occurs, it can function as a pump wave for a second Brillouin wave and so on. In contrast, single-line operation can be obtained by setting waveguide loading to a value that critically couples the pump wave at the target laser output power. This operational mode suppresses oscillation on all modes leading to high side-mode suppression. Figure ~\ref{fig:LLcurve} shows the output power versus coupled input power for single line operation. The lower right inset to the figure is the optical spectrum showing both the Stokes line and the suppressed pump wave.  The threshold for the device shown is 60 microWatts, and the differential pumping efficiency is 95\% . Side mode suppression of the neighboring cavity modes exceeds 70 dB at around 1 milliWatt output power (measured by observing the beatnote of the main mode and weak side modes on an electrical spectrum analyzer).

The threshold for SBL action is given by the following expression.
\begin{equation}
\label{eq:threshold}
P_{th}=\frac{2\pi^{2}n^{2}V_{eff}}{g_{b}Q_{p}Q_{b}\lambda_{p}\lambda_{b}}
\end{equation}

\begin{figure}[h]
\begin{center}
\scalebox{0.46}{\includegraphics{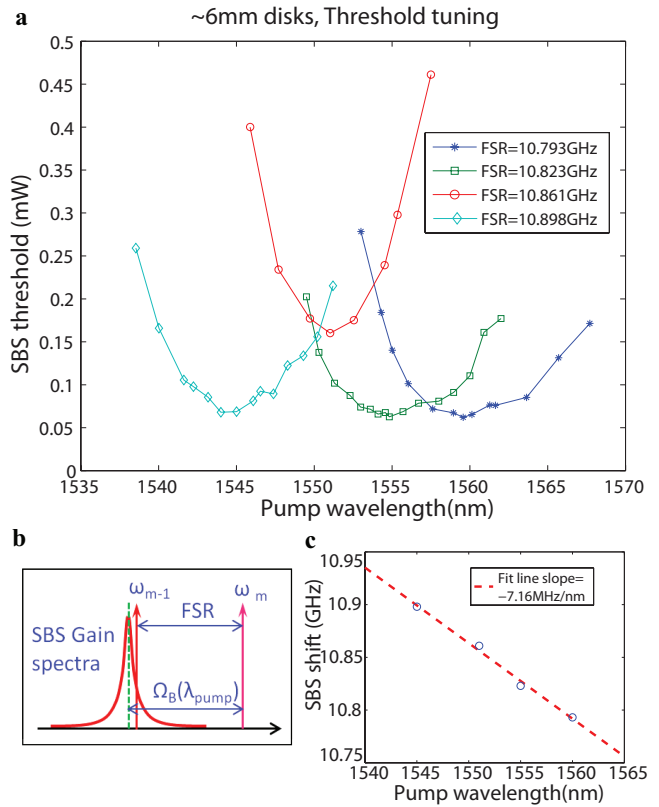}}
\caption{\label{fig:threshold}\textbf{Illustration of tuning control of the SBL devices. a}, SBL threshold is measured as a function of pump wavelength using four slightly different resonator diameters. FSRs are indicated in the legend. \textbf{b}, Illustration of the control of the SBS gain with the change of FSR and pump wavelength. \textbf{c}, Measured Brillouin-shift frequency (circles) versus pump wavelength.}
\end{center}
\end{figure}

{\noindent}Beyond the importance of high cavity Q factor evident in this expression, it is essential to maintain a large SBL gain parameter, $g_{b}(\Delta\omega - \Omega_{B})$ (where gain = $g_{b}P_{pump}$, $\Delta\omega$ is the free-spectral range, and $\Omega_{B}$ is the Brillouin shift).  Because the gain spectrum is relatively narrow (typical full-width half maximum is 20-60 MHz \cite{SBSgain,fiberSBS1}), this requires a precise match of the free-spectral-range to the Brillouin shift. $\Omega_{B}$ depends on the pump wavelength $\lambda_{p}$ and phonon velocity $V_{a}$ through the relation  $\Omega_{B}/2\pi=2nV_{a}/\lambda_{p}$. An illustration of the control possible using the new resonator geometry is provided in Figure \ref{fig:threshold}a where four devices having diameters of 6020, 6044, 6062 and 6080 microns (lithography mask size) were tested at a series of pump wavelengths in the 1500 nm band.
In each device, the pump wavelength was sequentially tuned along resonances belonging to the same azimuthal mode family. 
 The minimum threshold for each device corresponds to excitation at the Brillouin gain maximum  (i.e., $g_{b}(\Delta\omega - \Omega_{B}=0)$).  The rise in threshold away from the minimum (for a given resonator diameter) reflects tuning of the Brillouin shift frequency with pump wavelength noted above.  The variation of this peak frequency shift versus the corresponding pump wavelength is plotted in Figure \ref{fig:threshold}c. There is good agreement with the expected theoretical dependence. 

To characterize the SBL frequency noise, a Mach-Zehnder interferometer having a free spectral range of 6.72 MHz is used as a discriminator and the transmitted optical power is detected and measured using an electrical spectrum analyzer (ESA). To suppress the intensity noise, the complementary outputs of the interferometer were detected using a balanced receiver. This ESA spectrum is related to the frequency-fluctuation spectral density, $S_{\nu}(f)$, through the following relation:

\begin{equation}
\label{eq:MZI}
W_{ESA}(f)=\frac{V_{pp}^{2} 2 \pi^{2} \tau_{d}^{2} \textnormal{sinc}^{2}(\tau_{d} f) S_{\nu}(f)}{R_{L}}
\end{equation}

\begin{figure}[h]
\begin{center}
\scalebox{0.425}{\includegraphics{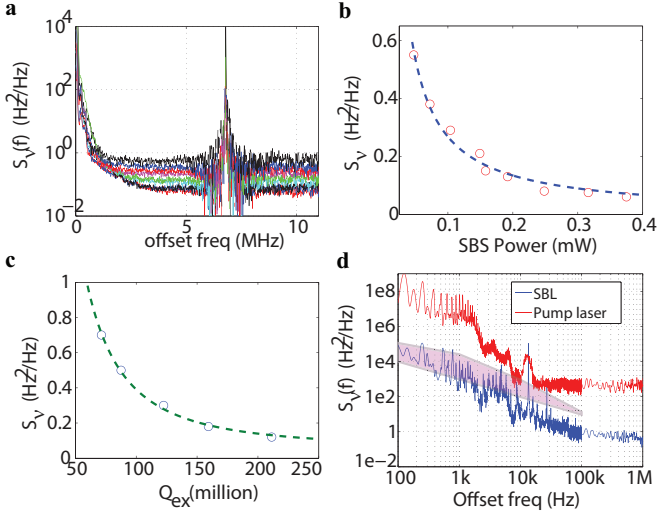}}
\caption{\label{fig:STnoise} \textbf{Measurements of the SBL frequency noise characteristics. a}, Laser frequency noise spectrum  at different output power levels from 0.047mW to 0.375mW (indicated by color). \textbf{b}, ST noise 
plotted versus  output power. The dashed line is an inverse power fit to the data. \textbf{c}, ST noise plotted versus the  external cavity quality factor $Q_{ex}$. The dashed line is a fit using the function $\frac{Q_{0}+Q_{ex}}{Q_{ex}^{2}}$. \textbf{d}, A typical SBL frequency noise spectrum  and the pump laser (external cavity diode laser) frequency noise spectrum at offset frequencies from 100Hz to 1MHz. The shaded region is the frequency noise performance of commercial, narrow linewidth fiber lasers.}
\end{center}
\end{figure}

{\noindent}where $\tau_{d}$ is the Mach-Zehnder delay and $V_{pp}$ is the peak-to-peak voltage of the detected MZI output over one fringe. Using this formula, the frequency-fluctuation spectral density is plotted in Figure \ref{fig:STnoise}a. The singularity in the plots at 6.72 MHz is an artifact of the data conversion near the zero of the sinc$^{2}$ function. The frequency-fluctuation spectra have a relatively flat (white noise) region for carrier offset frequencies above 2 MHz and then a 1/f-like region at frequencies below 2 MHz.  The value of the white noise region is plotted both as a function of SBL power and external cavity Q factor in Figures \ref{fig:STnoise}b and \ref{fig:STnoise}c, respectively. Also plotted are fits to inverse power and inverse $Q^{2}$ curves. These dependences as well as calculation confirm that the measured frequency noise component is the ST quantum noise of the laser. The minimum value of 60 milliHz$^2$/Hz is to our knowledge the smallest recorded ST noise for any chip-based laser. It is also interesting to note that, to the authors knowledge, the ST noise dependence on loading has not previously  been recorded. This, normally difficult measurement, was possible here on account of the ability to vary the taper loading of the resonator \cite{ideality}.  

The 1/f noise that appears at lower carrier offset frequencies is given in Figure \ref{fig:STnoise}d and approximately tracks a similar-shaped noise spectrum in the laser pump over this frequency range. However, the absolute level of 1/f noise of the SBL at a given offset frequency is reduced by about 30dB relative to the 1/f noise in the pump.  Indeed, the level of technical noise in this band is comparable to  several commercial fiber lasers that were characterized as part of this study. As such, the performance of the SBL, both in the quantum limited ST regime and the technical-noise limited 1/f regime, is excellent.

The ability to obtain frequency noise levels comparable to fiber lasers, but for on-chip operation opens new possibilities for miniaturization and integration in coherent communication, high-stability microwave oscillators, and potentially in remote sensing. In each of these cases, the SBL device provides a new role as a spectral purifier, boosting the coherence of the pump wave by several orders of magnitude. The relatively small frequency shift created in this process (about 11 GHz) is easily compensated in the pump. For example, low coherence DFB lasers are manufactured with wavelengths set on the ITU grid by control of an integrated grating pitch. However, final control is provided by temperature tuning of the fully packaged device. A DFB laser could be tuned through this same process to function as an SBL pump so that the emitted SBL wavelength resides at the desired ITU channel. In this way, the existing WDM infrastructure could be adapted for high coherence operation in optical QAM systems. The frequency noise levels demonstrated here exceed even state of the art monolithic semiconductor laser by 40dB. Using the measured phase noise, it is estimated that Square 1024-QAM formats could be implemented using an SBL generated optical carrier at 40GB/s.  

While the current devices use a taper coupling for launch of the pump and collection of laser signal, the ability to precisely control the resonator boundary enables use of microfabricated waveguides for this process. Several designs are under investigation, the implementation of which will extend the range of applications of the SBL devices.  For example, the SBLs demonstrated here are candidates for locking to a reference cavity so as to create Hertz or lower long-term linewidths.  Such a source on a chip might one day be combined with microcomb technology to realize a compact and high-performance microwave oscillator. At the table-top scale, these comb-based systems have recently exceeded the performance of cryogenic electronic oscillators \cite{Diddams}.

%%%%%%%%%%%%%%%%%%%%%%%%%%%%%%%%%%%%%%%%%%%%%%%%%%%%%%%%%%%%%%%%%%%%%%%%%%%%
%%%%%%%%%%%%%%% Acknowledgments, Figures, and References Section %%%%%%%%%%%
%%%%%%%%%%%%%%%%%%%%%%%%%%%%%%%%%%%%%%%%%%%%%%%%%%%%%%%%%%%%%%%%%%%%%%%%%%%%
\hspace{10 mm}

\textbf{Acknowledgments}
The authors would like to thank Scott Papp and Scott Diddams for helpful comments. We gratefully acknowledge the Defense Advanced Research Projects Agency under the Orchid programs and also the Kavli Nanoscience Institute at Caltech. 

\hspace{10 mm}

% Set the ending of a LaTeX document
\end{document}